\documentclass[useAMS]{mn2e}
\usepackage{times}
\usepackage{psfig}
\def\be{\begin{equation}}
\def\ee{\end{equation}}
\begin{document}

\title[dense matter EoS in pulsar astrophysics .....]{Importance of constraining the dense matter Equation of State in pulsar astrophysics}

\author[Bagchi ]
{\parbox[t]{\textwidth}{
        Manjari Bagchi}\\
\vspace*{3pt} \\
\\  Tata Institute of Fundamental Research, Homi Bhaba Road, Colaba, Mumbai 400 005, India. }

\maketitle

\maketitle

\begin{abstract}

{We study the dependence of the surface magnetic fields of radio pulsars on the choice of Equations of State, pulsar masses and the values of the angle between the magnetic axis and the spin axis of the pulsars within simple dipole model. We show that the values of the surface magnetic field can be even order of magnitude different from its canonical values. This difference will effect any magnetosphere related model to explain observational features of radio pulsars and magnetars. We find a significant difference of the value of the surface magnetic field from the commonly quoted value for the faster member of the double pulsar system, $i.e.$ PSR J0737-3039A as here both the mass of the pulsar and the angle between the magnetic axis and the spin axis are known. Our study reveals the importance of constraining the dense matter Equations of State in pulsar astrophysics as well as hints an alternative way to constrain these by independent determination of the pulsar magnetic field. }

\end{abstract}


\section{Introduction}
\label{sec:intro}

Magnetized rotating neutron stars can be categorized into different categories based on their emission properties, like AXPs, SGRs, radio pulsars, RRATS {\it etc}. Although almost every day some new interesting discoveries relating these objects are being made using the advanced technologies, it is unfortunate that the true description of the dense matter constituting these objects is still not well understood. A number of Equations of State (EsoS) are available in the literature trying to describe the state of the matter in such extreme condition as inside neutron stars, see Lattimer \& Prakash (2007) for a few examples of the EsoS. Better knowledge about the dense matter EsoS will in the long run help us to understand observational features of neutron stars in a better extent.

Presently there are a number of approaches trying to constrain the dense matter EsoS through astronomical observations of compact stars. The usual approach is to determine the mass and the radius of the stars with the help of various observational features like gravitational redshifts ($z$) from spectral lines, cooling characteristics, kHz quasi-periodic oscillations (QPO) {\it etc.} (Lattimer \& Prakash 2007, Li $et~al.$ 1999, \"Ozel 2006, Zhang $et~al.$ 2007). But these methods are not foolproof, $e.g.$ the value of $z$ used in \"Ozel's analysis of EXO 0748$-$676 can not be reproduced as mentioned by Klahn $et~al.$ (2006). Moreover, to constrain EsoS from QPO observations, one need to believe in some specific model of QPO which is again a subject of debate. Another alternative method might be the measurement of the moment of inertia from the faster component (A) of the double pulsar system PSR J0737-3039A/B (Lattimer \& Schutz 2005, Bagchi $et~al.$ 2009). As some high mass stars like PSR J1903+0327, EXO 0748$-$676 $etc.$ prefer neutron star models and some other stars like 4U 1728-34 (Li $et~al.$ 1999), prefer strange star models, it is possible that both family coexist in nature. But even then, we need some constrains as there are a number of EsoS for neutron stars and also for strange stars.

Because of the lack of any strong constrains on the dense matter EsoS till date, it is interesting to study the effect different choice of EsoS on the stellar properties. This is our motivation of studying the effect of EsoS in the determination of the values of the pulsar magnetic field and consequent results. Here we have chosen five EsoS, three for the neutron matter and two for the strange quark matter. We discuss the model of the pulsar magnetic field in section \ref{sec:mag_field} in short and then display our results in section \ref{sec:results}. Discussions and conclusions are given in section \ref{sec:disc} and section \ref{sec:concl} respectively.

\section{Magnetic field}
\label{sec:mag_field}

For a radio pulsar, the value of the surface magnetic field ($B_s$) is estimated by equating the spin down luminosity with the dipole radiation power and one gets the following expression :

\begin{equation}
B_s~=~\sqrt{\frac{3c^3}{8 \pi^2}\frac{I}{R^6 sin^2 \alpha}P\dot{P}}~=~B_{fac}~\sqrt{P\dot{P}}
\label{eq:def_mag}
\end{equation}
where $I$ is the moment of inertia of the pulsar, $R$ is its radius, $P$ is the spin period, $\dot{P}$ is the rate of change of the spin period, $\alpha$ is the angle between the magnetic axis and the spin axis of the pulsar and $c$ is the speed of light. $B_{fac}$ is defined as $\sqrt{\frac{3c^3}{8 \pi^2}\frac{I}{R^6 sin^2 \alpha}}$.

Due to the lack of knowledge about the actual value of $\alpha$, it is usually assumed to be  $90^{~\circ}$. Moreover, $I$ is taken as $10^{45}~ \rm{gm~ cm^2}$, $R$ as $10~\rm{km}$. Then $B_{fac}$  becomes $3.2 \times 10^{19}~ \rm { gm^{1/2} cm^{-1/2} sec^ {-3/2}}$. So Eqn. (\ref{eq:def_mag}) can be rewritten as :

\begin{equation}
B_{s}~=~3.2 \times 10^{19}\sqrt{P\dot{P}} ~~~G
\label{eq:conv_mag}
\end{equation}

\noindent Eqn. (\ref{eq:conv_mag}) is commonly used to calculate $B_s$ by putting the values of $P$ in seconds and $\dot{P}$ as dimensionless.

The problem with the above simplification is that, for a fixed value of the pulsar mass ($M$), one gets widely different values of $R$ and $I$ by using different EsoS for the matter. Moreover all the pulsars do not have the same value of $M$. In short, $R$ and $I$ can be different from their canonical values $10~\rm{km}$ and $10^{45}~ \rm{gm~ cm^2}$ respectively depending upon $M$ and the choice of the EoS. $\alpha$ can also have any value between $0^{~\circ}$ and $90^{~\circ}$. As a result the value of $B_{fac}$ will be different from $3.2 \times 10^{19}~ \rm { gm^{1/2} cm^{-1/2} sec^ {-3/2}}$. For this reason, it is worthwhile to study the dependence of the value of $B_s$ on $M$, $\alpha$ and EsoS. In the next section, we study this dependence using Eqn. (\ref{eq:def_mag}).

\section{Results}
\label{sec:results}

\begin{figure}
\centerline{\psfig{figure=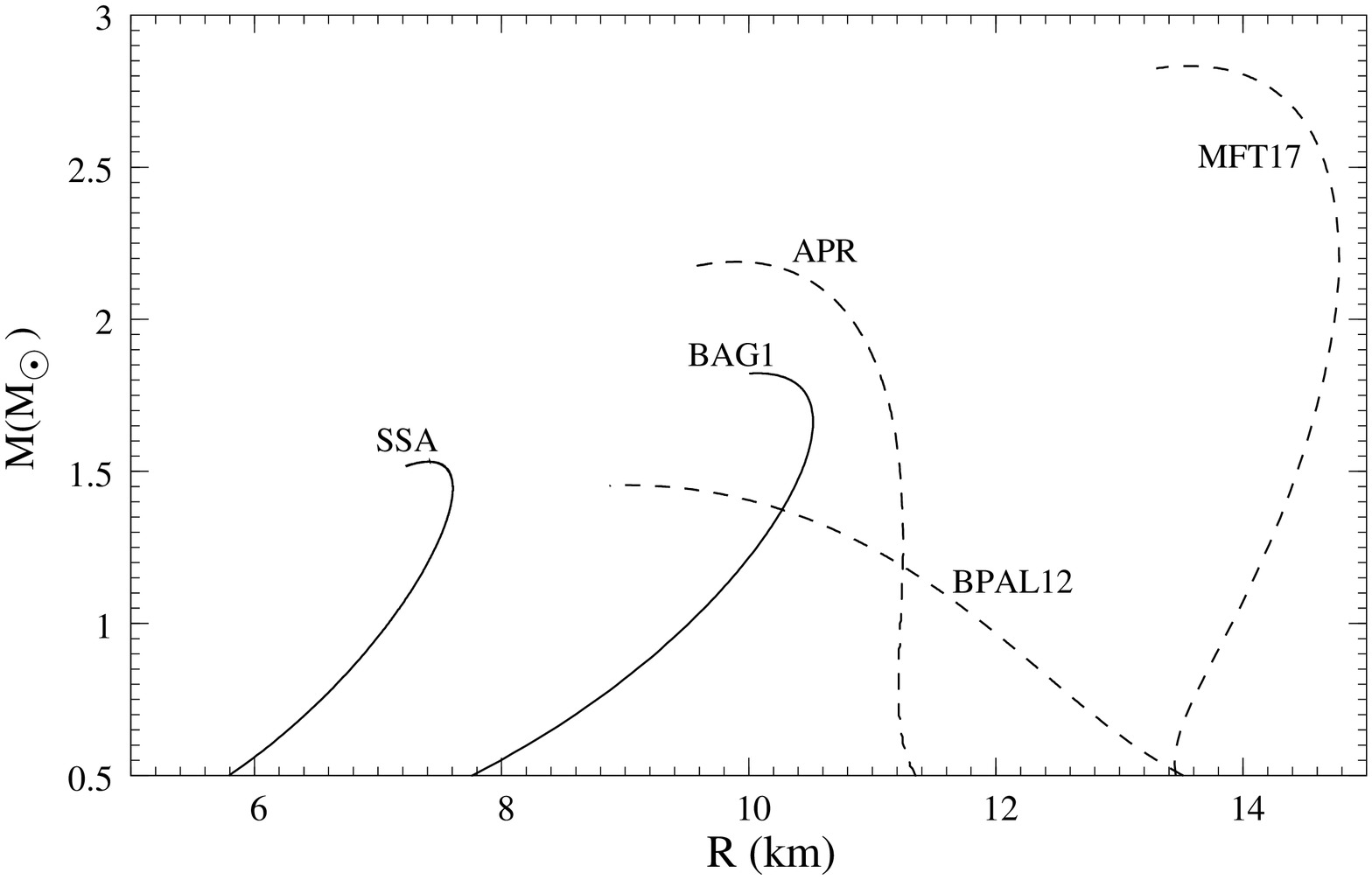,width=8cm}}
\centerline{\psfig{figure=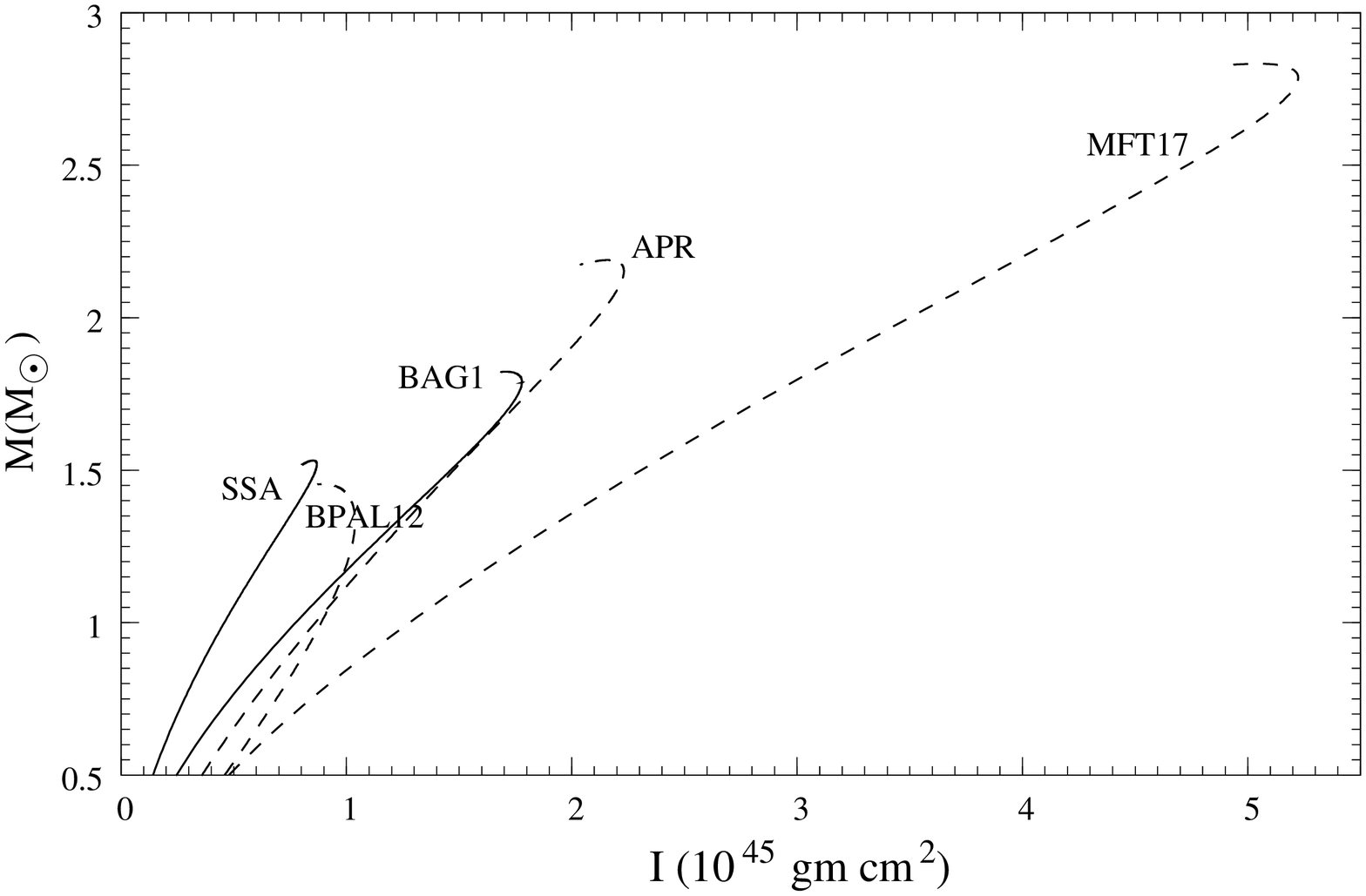,width=8cm}}
\caption{Panel (a) shows the mass - radius curves for various EsoS used in this work. Solid lines are for the strange quark matter EsoS whereas dashed lines are for the neutron matter EsoS. Panel (b) shows the moment of inertia - radius curves for the same EsoS.}
 \label{fig:mass_all_eos}
\end{figure}

Among numerous EsoS for the dense matter, we have chosen five - three for the neutron matter (namely EoS BPAL12, EoS APR and EoS MFT17 which are also used by Bejger $et~al.$ 2005) and two for the strange quark matter. One in the latter group is is EoS A (SSA) of Bagchi $et~al.$ (2006) and the other one is from Bag model (BAG1) with model parameters as : $B = 60.0~{\rm MeV / fm^3},~m_s = 150.0~ {\rm MeV},~ m_u = m_d = 0,~\alpha_c = 0.17$ where $B$ is the Bag parameter, $m_s$, $m_u$, $m_d$ are masses of s, u and d quarks respectively. These five EsoS are of widely different stiffness and so they give significantly different plots in the $M - R$ plane. In Fig. \ref{fig:mass_all_eos}, we show the $M - R$ and $M - I$ plots for all of these EsoS. It is worthwhile to mention here that MFT17 strongly overestimate and BPAL12 strongly underestimate the nuclear matter incompressibility. APR seems to be a better description of the nuclear matter. 

We have calculated the moment of inertia using the formalism derived by Kalogera $et~al.$ (1999) within Hartle's approximations.  We have found that even for an angular rotational frequency ($\Omega$) as high as 5000 $\rm sec^{-1}$, $B_{fac}$ obtained by this method differs maximum upto  $ \sim 10 \%$  from that obtained by using proper codes for rotating stars (RNS code, written by Nikolaos Stergioulas and available at http://www.gravity.phys.uwm.edu/rns/) for all of the EsoS mentioned earlier; the lower is the value of $\Omega$, the smaller is the difference. So for radio pulsars, the use of Hartle's approximations is well justified as the fastest pulsar yet discovered is PSR J1748 $-$ 2446{\it ad} with $\Omega~=~ 4500~\rm sec^{-1}$ (see ATNF pulsar catalog at http://www.atnf.csiro.au/research/pulsar/psrcat/).

\begin{figure}
\centerline{\psfig{figure=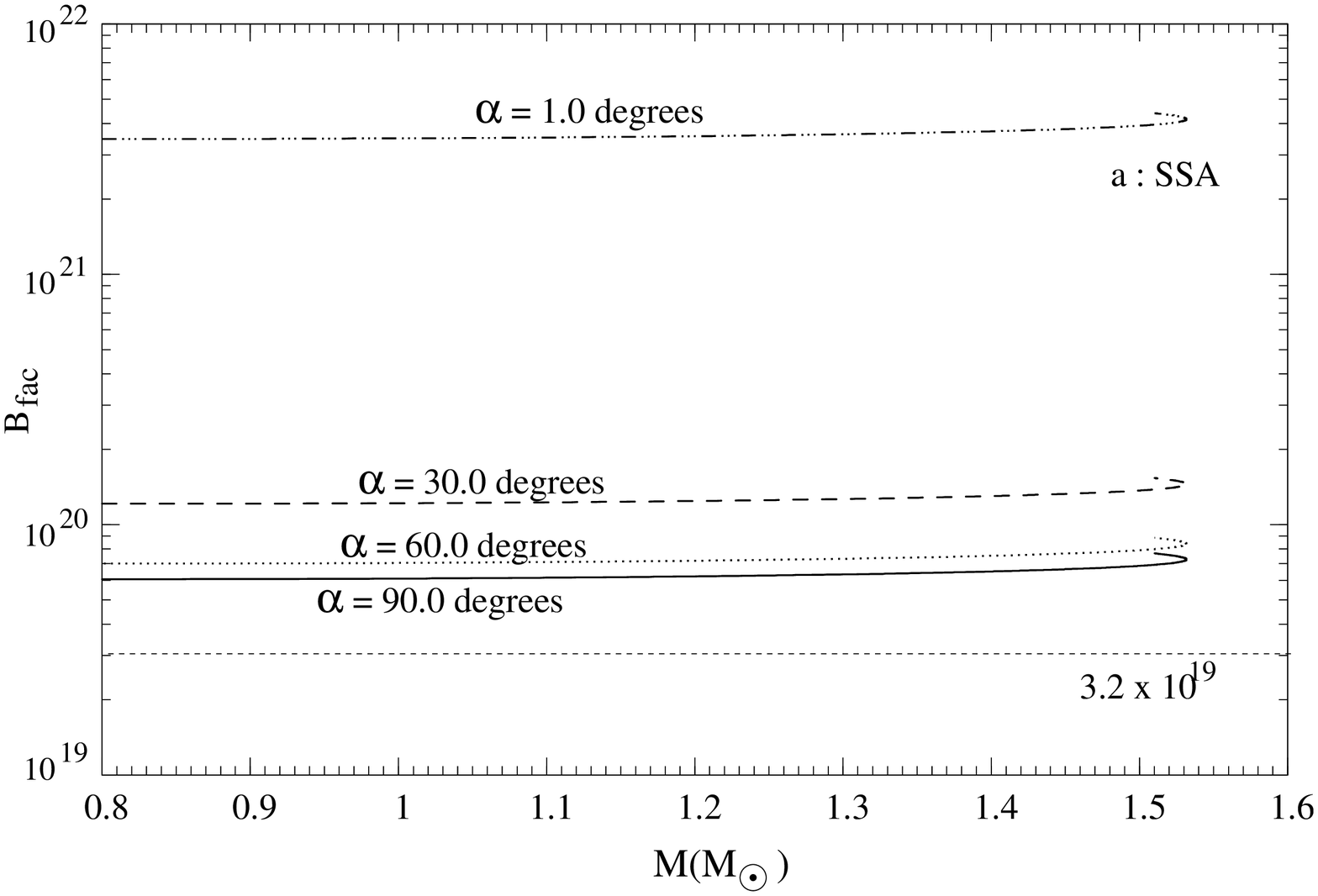,width=8cm}}
\centerline{\psfig{figure=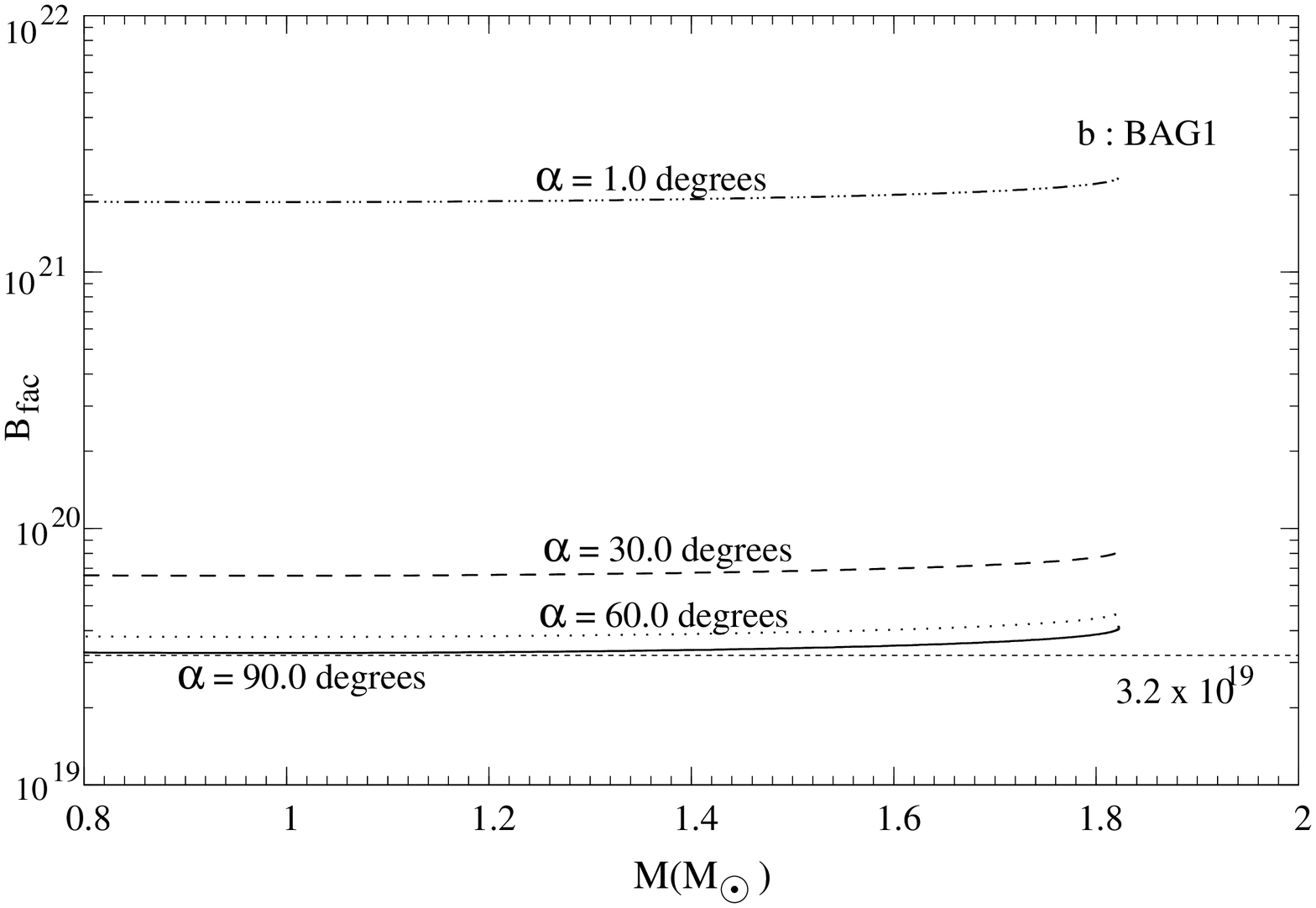,width=8cm}}
\caption{Variation of $B_{fac}$ with $M$ and $\alpha$ for different strange quark matter EsoS.  The unit of $B_{fac}$ is $ \rm {gm^{1/2} cm^{-1/2} sec^ {-3/2}}$. }
 \label{fig:bfac_all_alpha_strange}
\end{figure}

\begin{figure}
\centerline{\psfig{figure=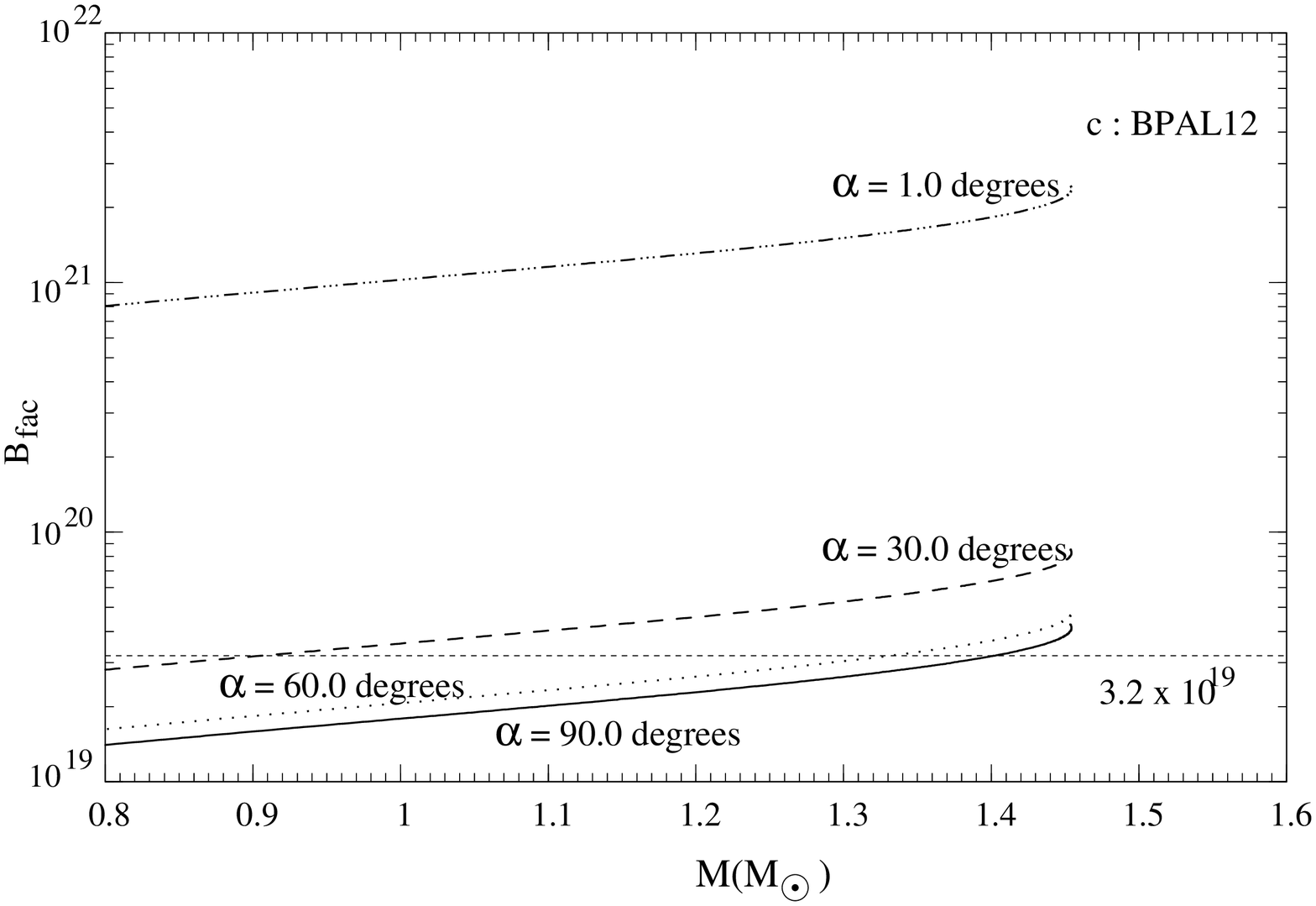,width=8cm}}
\centerline{\psfig{figure=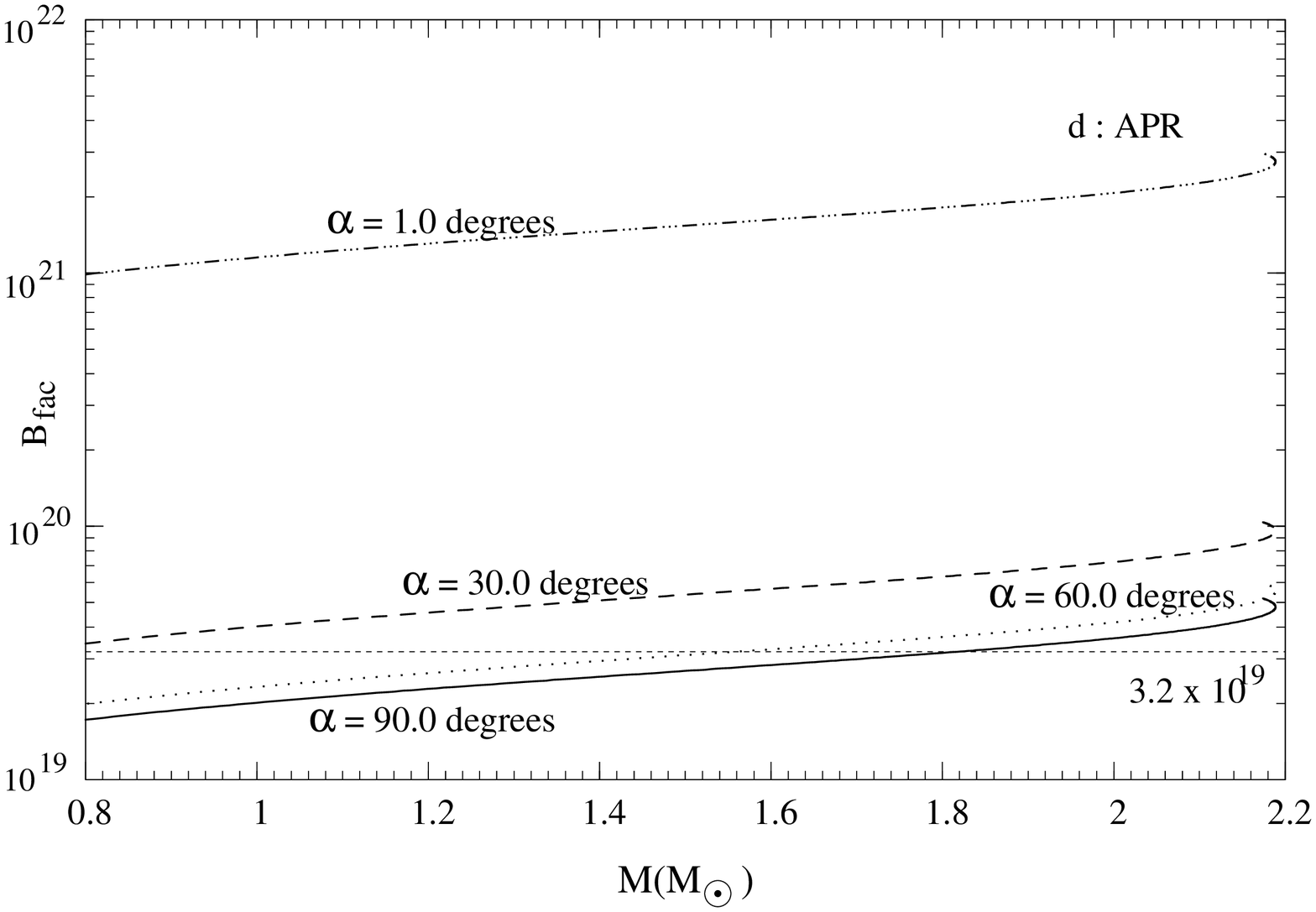,width=8cm}}
\centerline{\psfig{figure=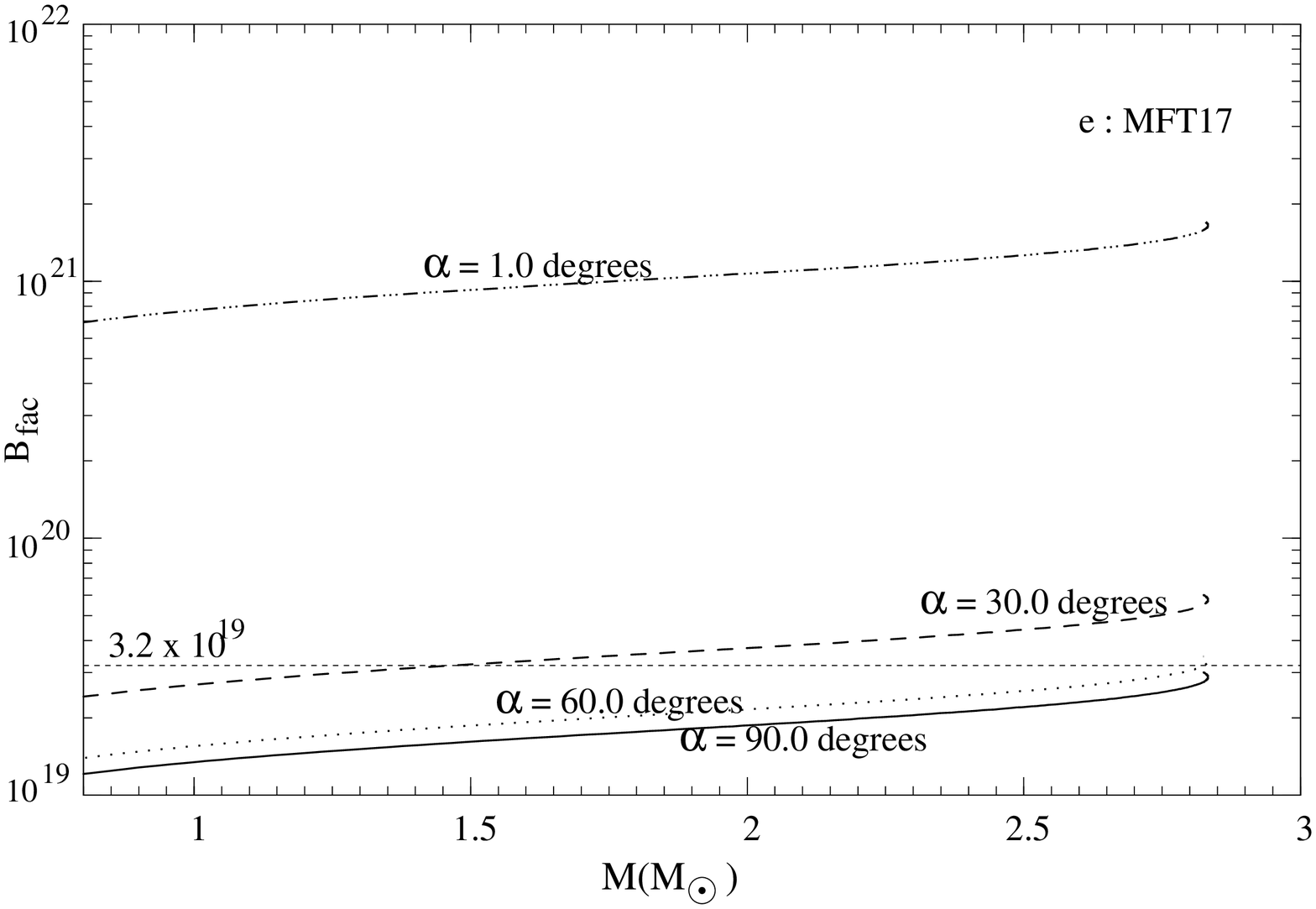,width=8cm}}
\caption{Variation of $B_{fac}$ with $M$ and $\alpha$ for different neutron matter EsoS.  The unit of $B_{fac}$ is $ \rm {gm^{1/2} cm^{-1/2} sec^ {-3/2}}$. }
 \label{fig:bfac_all_alpha_neutron}
\end{figure}

We have plotted the variation of $B_{fac}$ with pulsar masses  for different  EsoS (Figs. \ref{fig:bfac_all_alpha_strange} and \ref{fig:bfac_all_alpha_neutron}) and different values of $\alpha$ ($1^{\circ}$, $30^{\circ}$, $60^{\circ}$ and $90^{\circ}$). We find that for the strange quark matter EsoS, the value of $B_{fac}$ is usually large. For EoS SSA, the value of $B_{fac}$ is greater than $3.2 \times 10^{19}$ for any values of $\alpha$ and $M$. For EoS BAG1, $B_{fac}$ is greater than $3.2 \times 10^{19}$ for  $\alpha<90^{\circ}$, at $\alpha=90^{\circ}$, $B_{fac} \simeq 3.2 \times 10^{19}$. For the neutron matter EsoS, $B_{fac}$ is greater than $3.2 \times 10^{19}$ for very low values of $\alpha$, $e.g.$ $\alpha=1^{\circ}$ (for any values of $M$). But for $\alpha \geqslant 30^{\circ}$, $B_{fac}$ can be smaller than $3.2 \times 10^{19}$ depending upon the values of $M$. For EoS BPAL12, at $\alpha=30^{\circ}$, $B_{fac}<3.2 \times 10^{19}$ if $M<0.9 M_{\odot}$, at $\alpha=60^{\circ}$, $B_{fac}<3.2 \times 10^{19}$ if $M<1.3 M_{\odot}$, at $\alpha=90^{\circ}$, $B_{fac}<3.2 \times 10^{19}$ if $M<1.4 M_{\odot}$. For EoS APR, at $\alpha=30^{\circ}$, $B_{fac} > 3.2 \times 10^{19}$ for any values of $M$, at $\alpha=60^{\circ}$, $B_{fac}<3.2 \times 10^{19}$ if $M<1.6 M_{\odot}$, at $\alpha=90^{\circ}$, $B_{fac}<3.2 \times 10^{19}$ if $M<1.8 M_{\odot}$. For EoS MFT17, at $\alpha=30^{\circ}$, $B_{fac}<3.2 \times 10^{19}$ if $M<1.5 M_{\odot}$, but for  $\alpha=60^{\circ}$ and $\alpha=90^{\circ}$, $B_{fac}<3.2 \times 10^{19}$ for any values of $M$.

For a fixed $M$ and fixed EoS, $B_{fac}$ is minimum ($B_{fac,~min}$) for $\alpha=90^{\circ}$, giving the minimum possible value of the $B_s$ ($B_{s,~min}$) for that $M$ and that EoS. From Figs. \ref{fig:bfac_all_alpha_strange} and \ref{fig:bfac_all_alpha_neutron}, it is clear that $B_{fac,~min}^{SSA} > B_{fac,~min}^{BAG1}>B_{fac,~min}^{BPAL12}>B_{fac,~min}^{APR}>B_{fac,~min}^{MFT17}$ for $M \leqslant 1.5~M_{\odot}$ giving $B_{s,~min}^{BAG1}>B_{s,~min}^{BPAL12}>B_{s,~min}^{APR}>B_{s,~min}^{MFT17}$.

Let $\Delta B_{s}\vert_{M} $ denote the change in $B_s$ for a change $\Delta \alpha$ in $\alpha$ keeping $M$ fixed and $\Delta B_{s}\vert_{\alpha} $ denote the change in $B_s$ for a change $\Delta M$ in $M$ keeping $\alpha$ fixed. Then for $M$ being in the range of $0.8 - 1.5~M_{\odot}$, $\Delta B_{s}\vert_{\alpha}^{BPAL12}> \Delta B_{s}\vert_{\alpha}^{SSA} >\Delta B_{s}\vert_{\alpha}^{APR}>\Delta B_{s}\vert_{\alpha}^{MFT17}>\Delta B_{s}\vert_{\alpha}^{BAG1} $ for equal values of $\Delta M$ for each EoS; $\Delta B_{s}\vert_{M}^{SSA} > \Delta B_{s}\vert_{M}^{BAG1}> \Delta B_{s}\vert_{M}^{BPAL12}> \Delta B_{s}\vert_{M}^{APR}>\Delta B_{s}\vert_{M}^{MFT17} $ for equal values of $\Delta \alpha$ for each EoS. 

Our study clearly reveals that in addition to constraining the EoS, the knowledge of $\alpha$ is also necessary to know the actual value of $B_s$ for a pulsar of known $M$, $P$ and $\dot{P}$. Sometimes it is possible to determine $\alpha$ observationally. As an example, for PSR J0737$-$3039A, Jenet \& Ransom (2004) have found $\alpha$ to be either $1.6^{~\circ} \pm 1.3^{~\circ}$ or $14^{~\circ} \pm 2^{~\circ}$ by careful study of the pulse profile while polarimetric study by Demorest $et~al.$ (2004) gives $\alpha~=~4^{~\circ} \pm 3^{~\circ}$. In table \ref{tb:j0737mag}, we present the values of $B_s$ for PSR J0737$-$3039A obtained from Eqn. (\ref{eq:def_mag}) with different EsoS. We find that $B_s$ can be even orders of magnitude higher than the canonical value obtained by using Eqn. (\ref{eq:conv_mag}).

\begin{table}
\caption{Surface magnetic field ($B_s$) for PSR J0737$-$3039A with different values of $\alpha$ as estimated by Jenet \& Ransom (2004) and Demorest $et~al.$ (2004). The canonical value of $B_s$ for this pulsar is $0.064 \times 10^{11} $ G (using $B_{fac} = 3.2 \times 10^{19} ~ \rm {gm^{1/2} cm^{-1/2} sec^ {-3/2}}$). Pulsar parameters are $M~=~1.3381~M_{\odot}$, $P~=~22.699~\rm{ms}$ and $\dot{P}~=~1.7599 \times 10^{-18}$, $\sqrt{P\dot{P}}~=~1.9987 \times 10^{-10}~\rm{sec^{1/2}}$. }
\begin{center}
\begin{tabular}{l c c c}
\hline
  &   $B_s$  &    &  \\ 
&   $ \times 10^{11}$ G &    & \\ \hline 
 EoS $\downarrow$ $\backslash$  $\alpha$ $\rightarrow$  & $1.6^{\circ}$  & $4.0^{\circ}$  & $14.0^{\circ}$ \\
\hline
 SSA   & 4.577  & 1.829   &  0.528  \\
BAG1   & 2.378 & 0.955 & 0.275  \\
BPAL12  & 1.999  &  0.803 &  0.232   \\
APR  &  1.763  &  0.706   &  0.204  \\
MFT17  & 1.099 & 0.440  & 0.126  \\ 
 \hline
\end{tabular}
\end{center}
\label{tb:j0737mag}
\end{table}

\section{Discussions}
\label{sec:disc}

We have seen that the actual value of $B_s$ can vary from its canonical value even orders of magnitude depending on the EoS, $M$ and $\alpha$. This difference may effect various calculations involving the value of $B_s$ like the study of Alfven QPOs in magnetars (Sotani, Kokkotas \& Stergioulas 2008), the study of timing properties of magnetars (\"Ozel 2002), the study of magnetorotational collapse (Lipunov \ Gorbovskoy 2008) and various observed parameters of magnetars (Malov \& Machabeli 2006) and so on.

In Table \ref{tb:j0737mag}, we showed the possible departure of $B_s$ of the faster member (A) of the double pulsar system PSR J0737-3039A/B from the canonical value. Presently we can not do the same for the slower member PSR B as for this source, the value of $\alpha$ is unknown. But from Fig \ref{fig:bfac_all_alpha_strange} and \ref{fig:bfac_all_alpha_neutron}, we can anticipate the value of $B_s$ to be significantly different from its canonical value. This fact will effect our understanding of the observed features of the double pulsar system in a great deal. As an example, once in a orbit,  PSR A gets eclipsed by the magnetosphere of PSR B (which has a larger magnetosphere due to the slower rate of rotation). Lyutikov \& Thompson (2005) modeled the eclipse as the attenuation of PSR A's radio beams by the synchrotron absorption in the closed magnetic field lines of PSR B. This model gives absorption coefficients as a function of the magnetic field of the pulsar B. Intensity and the polarization properties of the radio beam of PSR A during the eclipse is a function of the absorption coefficients and thus a function of the magnetic field of PSR B. So any study of the eclipse light curve or the polarization properties to extract pulsar parameters might be effected by the use of the canonical value of $B_{s}$ of PSR B.

There is a common practice of drawing {\it iso-}magnetic field lines in the  $P\dot{P}$ diagram for the pulsars using $B_{fac}=3.2 \times 10 ^{19}$. These line should be considered as {\it iso-}$B_{s,~min}$ lines (as $\alpha=90^{\circ}$), instead of {\it iso-}$B_{s}$ lines. Even then, only EoS BAG1 gives these values of $B_{s,~min}$ for $M=0.8-1.4 M_{\odot}$ as here $B_{fac,~min} \simeq 3.2 \times 10^{19}$. EoS SSA will give higher values; EsoS BPAL12, APR and MFT17 will give lower values in this mass range. Following the same logic, the lowest value of $B_{s,~min}$ observed so far (as mentioned in Ray Mandal, 2006; see their Fig 1) will shift to a higher value for EoS SSA and to some lower values for EsoS BPAL12, APR and MFT17 with $M=0.8-1.4 M_{\odot}$.  

Eqn. (\ref{eq:conv_mag}) is derived by equating the pulsar spin down radiation with the magnetic dipole radiation power. This model predicts the pulsar breaking index $n=\Omega \ddot{\Omega} / \Omega^{2}~=3$. For some pulsars, the values of $n$ have been measured observationally and most of these are found to be less than 3  $e.g.$ Crab pulsar has $n=2.515 \pm 0.005 $, PSR B1509$-$58 has $n=2.837 \pm0.001 $ and PSR B0540$-$69 has $n=1.81 \pm 0.07 $ (Lyne \& Smith 2006 and references there in). This fact hints to the necessity of an improved model for the spin down of pulsars. Xu \& Qiao (2001) proposed a model where both the dipole radiation and the longitudinal current outflow due to an unipolar generator take place. But the value of $\alpha$ they needed to fit the observed values of $n$ did not agree with observationally determined values of $\alpha$. So even a better model is needed to understand pulsar magnetosheric properties better which might give even further different values of $B_s$ than those obtained in this work. But the beauty of this work is that it reveals even orders of magnitude difference in the values of $B_s$ from its canonical values within the simple dipole model.

\section{Conclusion}
\label{sec:concl}

We have seen that depending upon EoS, $M$ and $\alpha$, the value of $B_s$ can differ significantly from the value quoted in the literature. The value of $B_s$ also determines the value of the magnetic field throughout the pulsar magnetosphere by the relation $B(r) \propto B_s /r^3$ where $r$ is the radial distance from the star surface. So correct understanding of the dense matter physics will help us to understand pulsar magnetosphere related phenomena better.

On the other hand, future technology might enable us to determine neutron star magnetic field strength ($B_s$) independently from the spectral analysis (like synchrotron absorption lines, Zeeman splitting etc). The values of $\alpha$, $M$, $P$ and $\dot{P}$ can be determined form timing analysis. Simultaneous determination of all these five quantities will provide an alternative way to constrain the dense matter EoS.

\end{document}